# *Preprint*

## Asphaltene Aggregation due to Waterflooding (A Molecular Dynamics Study)

**Salah Yaseen** [*]

Department of Chemical Engineering, University of Illinois at Chicago, Chicago, IL 60607-7052, USA; syasee3@uic.edu, salahyaseen1983@gmail.com

**G.Ali Mansoori**

Departments of Bio- and Chemical Engineering, University of Illinois at Chicago, Chicago, IL 60607-7052, USA; mansoori@uic.edu

[*] Corresponding author

## Abstract

In the present study, we report our findings on various asphaltenes-water interactions at a high reservoir condition (550 K and 200 bar) and the role of water in asphaltene association during the waterflooding process. Molecular dynamics (MD) simulations are performed on oil/water and oil/brine systems. The oil phase is composed of asphaltenes (10 wt%) and o.xylene in which asphaltenes are entirely soluble. Seven different model-asphaltenes with diverse molecular weights, architectures, and heteroatom contents are employed. Interestingly, MD results indicate that asphaltenes become less soluble in oil when water is partially misciblized in the oil phase. All model-asphaltenes containing nitrogen and/or oxygen atoms in their structures are prone to association. The driving force behind asphaltene aggregation is shown to be an asphaltene-water hydrogen bond. The utilization of the brine (3.5 wt% NaCl) instead of pure water induces a slight decrease in asphaltene propensity for aggregation due to salt-in effect. MD results suggest that face-to-face contact is the prominent stacking of asphaltene molecules in their aggregates.

## Nomenclature

| | |
|---|---|
| Ai | i=1, 2, ..., 6, notations for model-asphaltenes used in this study |
| GROMACS | GROningen MAchine for Chemical Simulations |
| HB | Hydrogen bonds |
| fs | Femto-second |
| MD | Molecular dynamics |
| NaCl | Sodium chloride |
| nm | Nano-meter |
| ns | Nano-second |
| O/B | Oil/brine system |
| O/W | Oil/water system |
| ps | Pico-second |
| NPT | Isothermal-isobaric ensemble (constant Number of molecules, Pressure, and Temperature) |
| o.xylene | Ortho-xylene |
| OPLS-AA | Optimized Potentials for Liquid Simulations-All Atom force field |
| RDF | Radial distribution function |
| SPC/E | Simple Point Charge/Extended, notation for a force field for water |

## Sub- and superscripts

| | |
|---|---|
| A | Asphaltene |
| Max | Maximum |
| W | Water |

## 1. Introduction

Petroleum asphaltenes are heavy and complex polycyclic organic components of crude oils. They may cause plugging of porous media of petroleum reservoirs, wells, production pipelines, and downstream facilities [Branco et al., 2001; Escobedo and Mansoori, 1997; Hu et al., 2004]. Recently, molecular dynamics (MD) simulations have been widely employed to investigate the behavior of asphaltene and other heavy organics in petroleum crudes [Khalaf and Mansoori, 2017; Pacheco- Sanchez et al., 2003]. We also reported results of our MD studies of the interactions between asphaltenes and their best and worst pure solvents, i.e., o.xylene and water. It was demonstrated that highly electronegative heteroatom segments of asphaltenes were responsible for the formation of asphaltene-water hydrogen bonds (HB$_{A-W}$). Also, it was found that both electrostatic and van der Waals interaction energies of asphaltene-water had significant contributions. In contrast, electrostatic interaction energy of asphaltene-o.xylene showed a minor contribution [Yaseen and Mansoori, 2017]. In here, we report our MD simulation results for the cases when water and brine are in contact with asphaltenic oil and water is partially miscible in oil at high pressures and temperatures.

Crude oil and water are co-produced during the secondary-recovery (waterflooding) processes. Some of the water initially exists in the reservoir rock pores within the hydrocarbon zone. Also, water is injected into a reservoir during the waterflooding process. Water injection is of considerable significance in petroleum production [Kim et al., 1990; Pacheco-Sanchez and Mansoori, 1997]. At high pressures and temperatures, water is partially miscible with petroleum fluid. In this case, studies into possible asphaltene-water interaction are of interest to establish the role of water in asphaltene deposition. Such essential studies are still in their infancy.

Asphaltene is partially dissolved and partially in colloidal, dispersed states in real crude oils [Kawanaka et al., 1989]. To establish the role of water in a de-stabilization of dissolved asphaltenes in oil, we choose o.xylene as the oil medium in which asphaltene is completely dissolved. O.xylene is the best hydrocarbon solvent for asphaltenes. In here, we report results of our investigation of the role of misciblized water and brine in oil containing only dissolved asphaltene, on the onset of destabilization of dissolved asphaltenes in oil. We speculate the onset results obtained here are identical with the case of dissolved asphaltenes in real crude oils.

## 2. MD Simulation Methodology

Seven model-asphaltenes identified as A1, A2, …, A7, are used in this study. Molecular structures of the seven model-asphaltenes are shown in Figure 1, and their properties are reported in the first four rows of Table 1. The selected model-asphaltenes cover a variety of asphaltene properties including molecular weight, architecture, and composition.

1. Molecular weights of asphaltenes in crude oils vary from several hundred to few thousands due to their association [Vazquez and Mansoori, 2000]. However, actual molecular weights of asphaltene monomers are in the range of a few hundred to about one thousand. In here, molecular weights of the selected model-asphaltenes are between 543 and 1197.

2. Molecular architecture of asphaltenes are classified into "island" and "archipelago" [Alvarez-Ramírez and Ruiz-Morales, 2013]. Island *a.k.a.* "continental" asphaltene consists of a highly condensed aromatic moiety and some side chains. Archipelago asphaltene is composed of multiple smaller condensed aromatic moieties connected by alkyl bridges. In this report, the island-type structure is represented by A1, A3, A4, A5, and A7, while A2 and A6 represent the archipelago-type structure.

3. Elemental analysis underlines that asphaltenes contain, in addition to carbon and hydrogen, amounts of sulfur, nitrogen, and oxygen [Escobedo and Mansoori, 1992]. In this report, the selected model-asphaltenes contain various heteroatom segments, which include amide, amine, carboxyl, hydroxyl, pyridyl, sulfide, thioanisole, and thiophene. Hydrogen atoms of asphaltenes that are bonded to the highly electronegative heteroatoms, such as oxygen and nitrogen, can donate hydrogen bonds [Yaseen and Mansoori, 2017].

At the start of every MD simulation, the simulation box is composed of two phases: the oil phase and the pure or saline water phase. The oil phase is chosen to be composed of o.xylene and asphaltene molecules. 24 asphaltene molecules (10 wt%) are first packed in the oil phase, and then, the simulation box is expanded in Y-direction and packed with pure or saline water. A brine phase is composed of water molecules and sodium and chloride ions. The concentration of sodium chloride in a brine phase is selected to be 3.5 wt%, consistent with the average ocean-waters salinity.

It is worthy to point out that the concentrations of model-asphaltenes and sodium chloride are maintained constant in all MD simulations. Also, both phases are constrained to have identical masses. Our MD simulations fall into two groups. The first group includes seven MD simulations, which are performed on an oil/water system. Similarly, the second group includes seven MD simulations that are performed on an oil/brine system.

GROMACS 5.1.2 software package is used. OPLS-AA force field is utilized to represent model-asphaltenes, o.xylene, and sodium chloride. Water is represented by SPC/E potential model. MD simulations are performed with periodic boundary conditions in $x$, $y$, and $z$ directions. Systems are energy-minimized using steepest descent method. They are performed for 20 ns (time step = 2 fs) in the isothermal-isobaric (NPT) ensemble. The temperature and pressure are kept constant at 550 K and 200 bar, and they adjust by a V-rescale thermostat (coupling constant = 0.1 ps) and the Parrinello-Rahman barostat (coupling constant = 2 ps). The standard velocity Verlet algorithm is used to integrate Newton's equations of motion.

All bond lengths are constrained with LINCS algorithm. The long-range electrostatic interactions are computed using a particle mesh Ewald method with a cut-off distance of 1.4 nm.

## 3. Simulation Results

### 3.1. Structural analysis of oil/brine and oil/water systems

Figure 2 represents snapshots of oil/water and oil/brine systems at the end of the MD simulations. Asphaltene molecules stay somewhere within the boundary of the oil phase. Some are suspended in the middle of the oil phase while others are located close to the interface. This behavior is caused by the strong van der Waals attraction forces between asphaltenes and o.xylene [Yaseen and Mansoori, 2017]. Sodium and chloride ions remain only in the brine phase. Elevated temperature causes disrupting of the mightily strong water-water hydrogen bonds ($HB_{W-W}$). Reduction of $HB_{W-W}$ allow some water molecules to diffuse into the oil phase. That is why o.xylene and water show partial miscibility in both phases.

### 3.2. Investigating the aggregation of model-asphaltenes via radial distribution functions

Radial distribution functions (RDFs) are a good measure of the state of asphaltene aggregation. The heights of RDF peaks of model-asphaltenes ($RDF^{Max}$) are computed and reported in Table 1. We are especially interested in RDFs peaks with a separation distance up to around 1.00 nm, as it considers the maximum spacing distance between the stacked asphaltene molecules. RDFs indicate that all model-asphaltenes except A5 and A6 exhibit a moderate to high tendency to aggregate. The maximum intensity of asphaltene aggregation is observed when A4 is used in oil/water system. RDFs of A5 and A6 indicate that these model-asphaltenes exhibit a very slight tendency to aggregate. The minimum level of asphaltene association is observed when A5 is used in oil/brine system.

It is important to point out here that the separation distance between the stacked asphaltene molecules depends mainly on the configuration of the aggregated asphaltenes [Gao et al., 2014]. It is believed that intermolecular stacking of asphaltene may fall into face-to-face, T-shaped, and parallel-displaced geometries [Pacheco-Sanchez et al., 2003]. The minimum separation distance is typically observed when asphaltene molecules are stacked face-to-face. Dickie and Yen model of asphaltenes indicated that the spacing distance between the aggregated face-to-face asphaltenes is 0.35-0.38 nm [Dickie and Yen, 1967]. However, other studies reported various separation distances up to around 0.5 nm [Frigerio and Molinari, 2011; Kuznicki et al., 2008]. In the case of parallel-displaced and T-shaped configuration, the spacing distance is assumed to be larger (up to around 1.00 nm), which depends on the size and structure of the asphaltene molecule [Headen et al., 2009]. From RDF data in Table 1, the minimum separation distances of A1, A2, A3, and A7 are 0.50, 0.46, 0.4, and 0.50 nm respectively. On the other hand, the minimum interlayer distance of A4 is 0.38 nm, which is in good agreement with the Dickie and Yen model of asphaltenes.

The configuration of stacked asphaltenes is strongly influenced by their structure [Pacheco-Sanchez et al., 2004] and the composition of oil bulk [Priyanto et al., 2001]. In Figure 3, configurational snapshots and RDFs are presented to study the structure of asphaltene aggregates. A1, A2, A3, A4, and A7 are selected because they show the highest aggregation intensity in this report as depicted earlier. Snapshots and RDFs of A1, A2, and A3 reveal that multiple stacking configurations exist in their aggregates. However, face-to-face stacking is the marked configuration between the stacked asphaltenes since more than 50% of molecules of every model-asphaltene are involved in this type of arrangement. For A4, the RDF shows a sharp peak at 0.38 nm, which implies that aggregates are almost stacked face-to-face. The presence of small peaks at 0.65, 0.79, and 1.00 nm indicates that A4 molecules are rarely stacking in parallel-displaced and T-shaped configurations. From the snapshot, the aggregated molecules of A4 are grouped in a rod-like fashion via face-to-face stacking. Snapshot and RDF of A7 underline that the probability of face-to-face, parallel-displaced, and T-shaped staking of A7 aggregates is somewhat comparable.

## 3.3. The distribution of asphaltene aggregates

The intensity of asphaltene aggregation is further quantified using the distribution of asphaltene aggregates. The average fraction of asphaltene in various association stages is calculated and reported in Table 1. It is evident that the number of aggregated asphaltene molecules in both oil/water and oil/brine systems is almost comparable. However, asphaltene molecules in oil/brine system tend to form smaller aggregate size than that in oil/water system. The largest aggregate observed is a heptadecamer, which is formed by A4 in oil/water system. All A4 molecules are involved in intermolecular contacts, and 71% of them exist in a single stacked structure. The distribution of asphaltene aggregates indicates that A1, A2, A3, and A7 are also prone to association. Among these model-asphaltenes, the higher rate of asphaltene aggregation is observed between A3 molecules. We notice that aggregates size of A1 and A7 is slightly higher than that of A2. On the other hands, the number of A1 and A7 monomers is fewer than the one of A2. These observations imply that the aggregation intensity of A1 and A7 is slightly higher than that of A2. Lastly, the most stable aggregate of A5 and A6 is a dimer. An abundance of A5 monomers and dimers implies that A5 is almost soluble in oil bulk.

### 3.4. The distribution of o.xylene and water molecules near the aromatic core of model-asphaltenes

To further quantify the impact of the miscibilzed water on the intensity of asphaltene aggregation, we investigate the distribution of o.xylene and water molecules around the aromatic core of model-asphaltenes. To fulfill this purpose, RDFs of asphaltene-o.xylene and asphaltene-water are computed and depicted in Figure 4. Since no sharp peaks are observed, it is concluded that neither o.xylene nor water molecules show a significant short-range structure near aromatic cores. RDFs of asphaltene-o.xylene indicate that the cut-off distance between the first layer of o.xylene molecules around the aromatic core of asphaltene is about 0.5 nm. This observation agrees with previous findings [Teklebrhan et al., 2016]. The heights of asphaltene-water RDFs are lower than that of asphaltene-o.xylene, which reveal that water exhibits less affinity for aromatic cores than o.xylene molecules. The suppression of the RDF of A4-o.xylene implies that there are few o.xylene molecules close to the aromatic core of A4. The removal of o.xylene molecules from the vicinity of the aromatic core suggests that the association of A4 is very noticeable compared to other asphaltenes. On the contrary, the A5-o.xylene RDF of implies that plenty of o.xylene molecules surrounds the aromatic core of A5. The abundance of o.xylene and the lack of water molecules near the aromatic core of A5 confirm the solubility of A5 by o.xylene molecules.

### 3.5. Hydrogen bonds of asphaltene-asphaltene and asphaltene-water

In addition to the primary role of $\pi$-$\pi$ interactions of aromatic moieties of asphaltenes (face-to-face stacking), it is well-established that asphaltene-asphaltene hydrogen bonds ($HB_{A-A}$) participate to some extent in asphaltene aggregation [Gray et al., 2011]. Also, it is evident that $HB_{A-A}$ may affect the shape of the aggregates. Depending on the location of highly electronegative heteroatoms, $HB_{A-A}$ may form side chain-side chain and/or side chain-aromatic core stacking. [Teklebrhan et al., 2016]. The number of statistically averaged $HB_{A-A}$, as well as $HB_{A-W}$, are computed using GROMACS built-in tools and reported in Table 1. Surprisingly, the contribution of $HB_{A-A}$ in the structure of asphaltene aggregates is very modest relative to those mentioned in previous studies [Sedghi et al., 2013; Takanohashi et al., 2003]. The employed model-asphaltenes have multiple highly electronegative heteroatoms. However, we identified as many as five $HB_{A-A}$ when A4 is used. The number of computed $HB_{A-A}$ is relatively tiny in the case of A1, A2, and A6. As expected, there is no $HB_{A-A}$ observed when A5 is used due to the absence of highly electronegative heteroatoms in its structure. Similarly, no $HB_{A-A}$ is noticed in case of A7 because this model-asphaltene has a pyridyl group only, which does not have hydrogen atom to donate hydrogen bond.

## 4. Discussion

Asphaltenes tend to be soluble in aromatic solvents such as benzene, toluene, and xylene. The ability of aromatic solvents to solubilize asphaltenes is attributed to $\pi$-$\pi$ interactions between solvent molecules and the aromatic core of asphaltenes. These non-covalent interactions would create a non-favorable environment for the association of asphaltene monomers [Khalaf and Mansoori, 2017; Mansoori, 1997; Pacheco-Sanchez and Mansoori, 1997]. In the present study, all model-asphaltenes except A5 and A6 tend to associate when water is partially miscible in the oil phase. These model-asphaltenes show various degrees of association. Their tendency to associate varies between a moderate, in the case of A1, A2, A3, and A7, and a strong aggregation, in the case of A4.

A5, which is almost soluble in the oil phase, does not have any polar heteroatoms segments in its structure. Contrarily, A1-A4, A6, and A7 share a similarity of containing polar heteroatoms segments in their structures. Asphaltene monomers containing polar atoms strongly interact with water molecules via hydrogen bonds ($HB_{A-W}$) (see Table 1). This interaction is attributed to the presence of plenty of water molecules in the oil phase. The accumulation of water molecules near polar atoms of model-asphaltenes considerably brings down the number of $HB_{A-A}$ and displaces o.xylene molecules. The nearest-neighbor layer of molecules that surrounds every asphaltene monomer is composed of o.xylene and water. Reduction of the o.xylene population near asphaltene monomers influences their ability to hinder the aggregation process.

The highest aggregation intensity is noticed when A4 is used (see Table 1). The structure of A4 includes an amide and two hydroxyl groups. An amide group serves as a lone-pair acceptor and donor simultaneously. The oxygen atom of carbonyl (C=O) may accept hydrogen bonds from water, and the hydrogen atoms of (N-H) may donate hydrogen bonds. $HB_{A-W}$ boosts the presence of water near A4 monomers and expels o.xylene molecules. So, the ability of o.xylene to serve as dispersant decreases. Additionally, A4 has a large and well-structured circle-like aromatic moiety, which is composed of 16 condensed aromatic rings. A strong $\pi$-$\pi$ interaction occurs between the sizeable aromatic core of A4 molecules. On the other hand, the misciblized water in the oil phase does not cause aggregation of A5 (see Table 1). This model-asphaltene does not attract water molecules due to the absence of highly electronegative heteroatoms. Also, it has several and long aliphatic side chains that hinder its tendency to form aggregates.

MD results indicate that A1, A3, and A7 exhibit a moderate tendency to form aggregates (see Table 1). Among this group of model-asphaltenes, the structure of A3 has the larger number of aliphatic side chains, which are known for their role in hindering asphaltene aggregation [Khalaf and Mansoori, 2017]. Also, A1 and A7 have well-constructed circle-like aromatic moieties with many fused aromatic rings, which induce the aggregation of asphaltenes [Pacheco-Sanchez et al., 2004]. However, A3 shows the higher levels of association. This may be attributed to the presence of two polar heteroatom segments in the structure of A3, namely carboxyl and hydroxyl, while each of A1 and A7 has a single polar heteroatom segment, which is carboxyl and pyridyl respectively. An average of two $HB_{A\text{-}W}$ per each A3 monomer are formed resulting in the accumulation of water molecules near polar heteroatom segments of A3. A1 and A7 form an average of 1.2 $HB_{A\text{-}W}$ per each asphaltene monomer, which are apparently lower than that of the $HB_{A\text{-}W}$ A3 case.

For archipelago model-asphaltenes, the association intensity of A2 is higher than the one of A6. That is because the structure of A2 includes a hydroxyl and two amine segments, while the structure of A6 includes a hydroxyl and pyridyl segments. Consequently, the number of $HB_{A\text{-}W}$ in the case of A2 is higher than that of A6 (see Table 1). Also, the structure of A2 has an aromatic moiety with seven fused aromatic rings, which is larger than the two aromatic moieties of A6.

Employment of brine, instead of pure water, causes a slight decrease in asphaltene aggregation levels. As an example, the RDF of A4 in an oil/brine system shows a lower peak height than that in an oil/water system by about 12%. This behavior is known as the "salt-in effect" [Wang et al., 2010]. This observation is similar to the computational finding of Jian et al., 2016. They investigated the behavior of asphaltene in toluene/brine (28 wt% NaCl) and toluene/water under normal conditions. They found that the replacement of pure water by brine brought about a slight decrease in the extent of asphaltene aggregation. Also, our MD results are in good agreements with the experimental observations of Shojaati et al., 2017 and Ameri et al., 2017, who investigated the effect of salinity on real crude oils at normal conditions.

# 5. Conclusions

In waterflooding process, brine is injected to displace crude oil from the reservoir porous media towards the production well to increase the production rates. Some water molecules are misciblized in the oil phase that may already contain asphaltenes along with other heavy organics like resins, petroleum waxes, diamondoids, *etc*. There is a great deal of literature dealing with the asphaltene-water interactions in normal conditions resulting in a stable emulsion. Notwithstanding our interest is the impact of the misciblized water on asphaltene aggregation under high reservoir conditions for which little or no data is available.

Our MD results indicate that asphaltene molecules remain somewhere in the oil phase at 550 K-200 bar. Furthermore, water and o.xylene molecules exhibit mutual miscibility. This report reveals that the misciblized water in oil phase causes aggregation of asphaltene molecules. The aggregation intensity of model-asphaltenes is quantified using radial distribution functions, hydrogen bonds and distribution of asphaltene aggregates. All model-asphaltenes except A5 and A6 show a considerable aggregation tendency as a result of water miscibility in oil phase. The aggregation intensity of these model-asphaltenes varies between moderate and strong association. Asphaltene-water hydrogen bonds are the driving force for asphaltene aggregation.

The attracted water molecules reduce the probability of o.xylene near asphaltene monomers and inhibit their ability to solubilize asphaltene molecules. Asphaltene-water interactions minimize asphaltene-asphaltene hydrogen bonds. Employment of brine instead of water enhances the solubility state of model-asphaltenes slightly due to salt-in effect. Face-to-face stacking is the preferred geometry in the asphaltene aggregate.

Acknowledgements

This research is supported, in part, by the Higher Committee for Education Development in Iraq (HCED), Office of the Prime Minister.

**Table 1.** Model-asphaltene properties and MD results in oil/brine (O/B) and oil/water (O/W) systems. All MD data are computed at 550 K and 200 bar.

| Property | | A1 | A2 | A3 | A4 | A5 | A6 | A7 |
|---|---|---|---|---|---|---|---|---|
| Origin [Ref] | | Athabasca [1] | Athabasca [1] | Athabasca [1] | Boscan [2] | Kuwait [3] | Khafji [4] | Maya [5] |
| Chemical Formula (MW) | | $C_{40}H_{30}O_2$ (543) | $C_{44}H_{40}N_2OS$ (645) | $C_{51}H_{40}O_3S_3$ (817) | $C_{85}H_{81}NO_3S$ (1197) | $C_{71}H_{96}S$ (982) | $C_{63}H_{69}NOS_2$ (920) | $H_{63}C_{57}N_1S_1$ (794) |
| Architecture | | Island | Archipelago | Island | Island | Island | Archipelago | Island |
| Hydrogen Bond Segments | | A Carboxyl | A Hydroxyl, An Amine | A Carboxyl, A Hydroxyl | An Amide, Two Hydroxyl | - | A Hydroxyl, A Pyridyl | Pyridyl |
| RDF^Max (r[nm]) of asphaltene peaks | | | | | | | | |
| O/B | 1st peak | 13.5 (0.50) | 16 .0(0.47) | 32.6 (0.40) | 193.0 (0.38) | 3.1 (0.43) | 7.9 (0.65) | 13.1 (0.51) |
| | 2nd " | 8.6 (0.75) | 4.2 (0.83) | 6.2 (0.84) | 29.0 (0.63) | - | - | 12.3 (0.85) |
| | 3rd " | - | - | - | 24.9 (0.79) | - | - | - |
| O/W | 1st " | 14.3 (0.50) | 16.9 (0.46) | 37.8 (0.40) | 223 (0.38) | 3.5 (0.43) | 8.7 (0.54) | 13.5 (0.50) |
| | 2nd " | 9.6 (0.73) | 5.9 (0.84) | 7.0 (0.82) | 31.2 (0.62) | - | - | 12.5 (0.86) |
| | 3rd " | - | - | - | 28.8 (0.79) | - | - | - |
| The average fraction of asphaltene in various association stages | | | | | | | | |
| O/B, O/W | Monomers | 0.13, 0.17 | 0.28, 0.25 | 0.13, 0.13 | 0.04, - | 0.50, 0.42 | 0.21, 0.21 | 0.17, 0.08 |
| | Dimers | 0.25, 0.08 | 0.17, 0.25 | 0.17, - | 0.08, - | 0.50, 0.58 | 0.67, 0.50 | 0.08, 0.25 |
| | Trimers | 0.25, 0.25 | 0.13, - | 0.13, 0.13 | -, 0.13 | - | 0.13, 0.13 | 0.25, 0.13 |
| | Tetramers | 0.17, 0.17 | - | 0.33, 0.17 | 0.17, 0.17 | - | -, 0.17 | - |
| | Pentamers | 0.21, - | 0.42, 0.21 | - | - | - | - | - |
| | Hexamers | - | - | 0.25, 0.25 | - | - | - | 0.50, 0.25 |
| | Heptamers | - | -, 0.29 | - | - | - | - | -, 0.29 |
| | Octamers | -, 0.33 | - | -, 0.33 | 0.33, - | - | - | - |
| | Nonamers | - | - | - | 0.38, - | - | - | - |
| | heptadecamers | - | - | - | -, 0.71 | - | - | - |
| The number of hydrogen bonds of asphaltene-asphaltene ($HB_{A\text{-}A}$) and asphaltene-water ($HB_{A\text{-}w}$) | | | | | | | | |
| O/B ($HB_{A\text{-}A}$, $HB_{A\text{-}w}$) | | 1, 29 | 1, 29 | 3, 47 | 5, 82 | 0, 0 | 1, 18 | 0, 5 |
| O/W ($HB_{A\text{-}A}$, $HB_{A\text{-}w}$) | | 1, 29 | 1, 29 | 3, 47 | 5, 83 | 0, 0 | 1, 18 | 0, 5 |

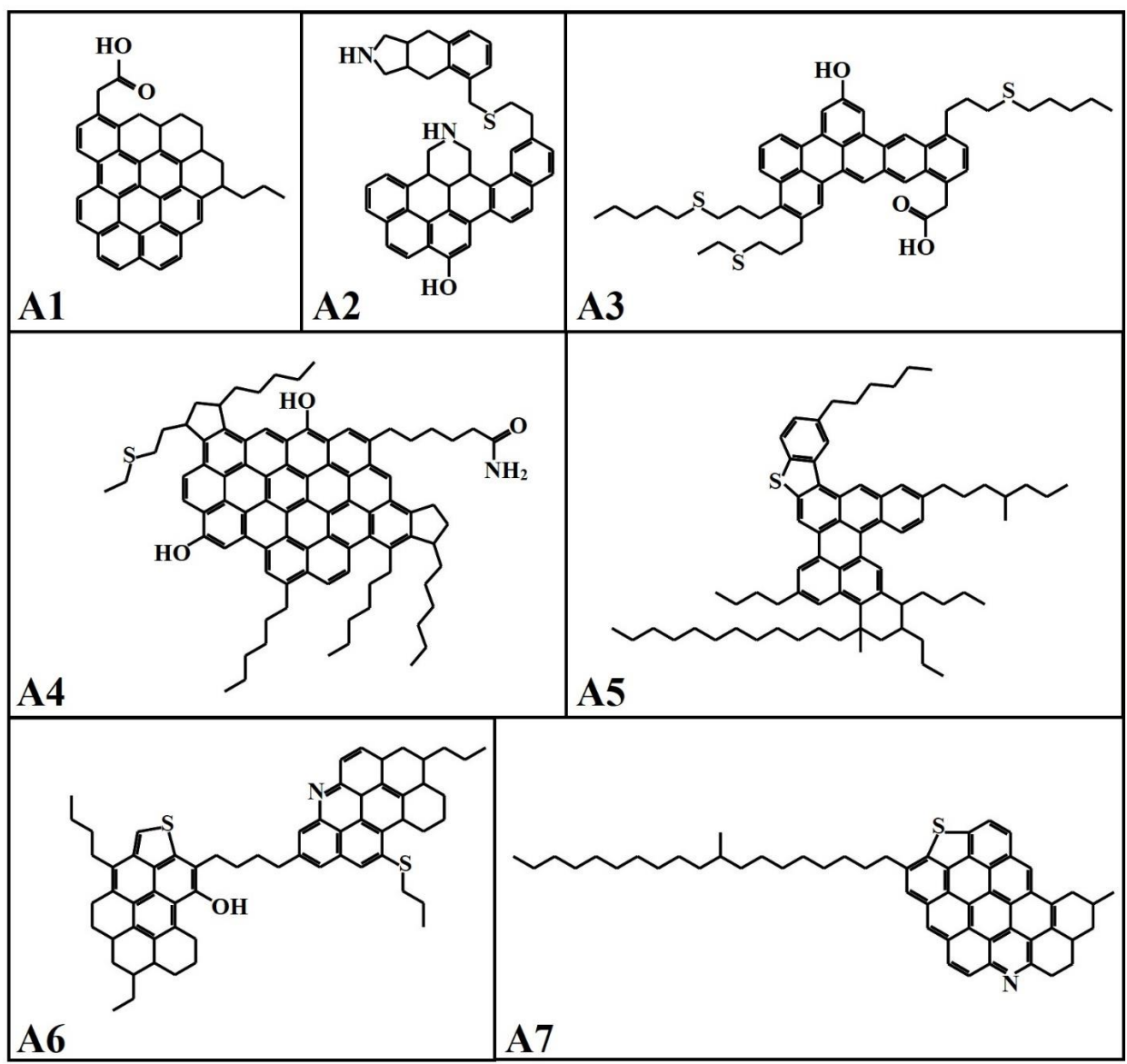

**Figure 1.** Structures of the seven model-asphaltenes used in this study.

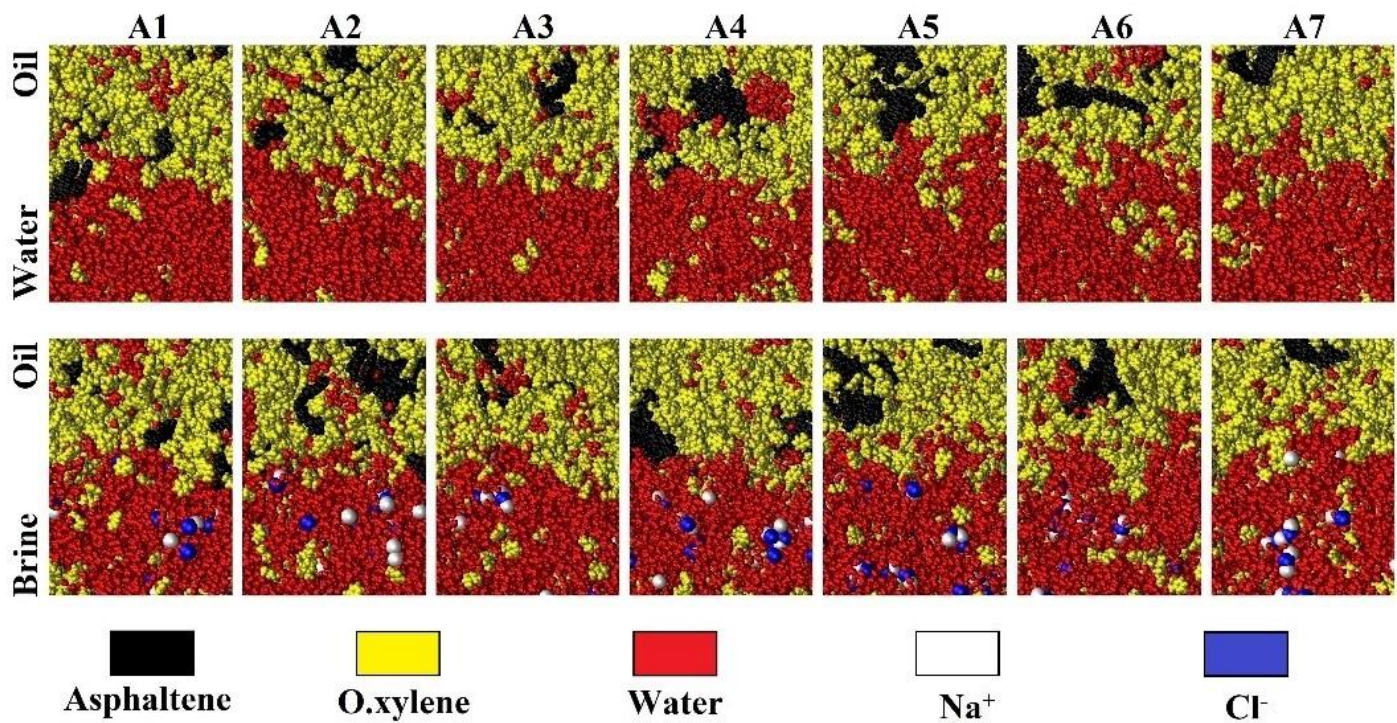

**Figure 2.** Front-view snapshots of simulation results at 550 K and 200 bar. The purpose of these snapshots is to show the mutual miscibility of water (red color) and o.xylene (yellow color) molecules. Also, these snapshots show that asphaltene molecules (black color) and ions (white and blue colors) are settled in oil and brine phases respectively.

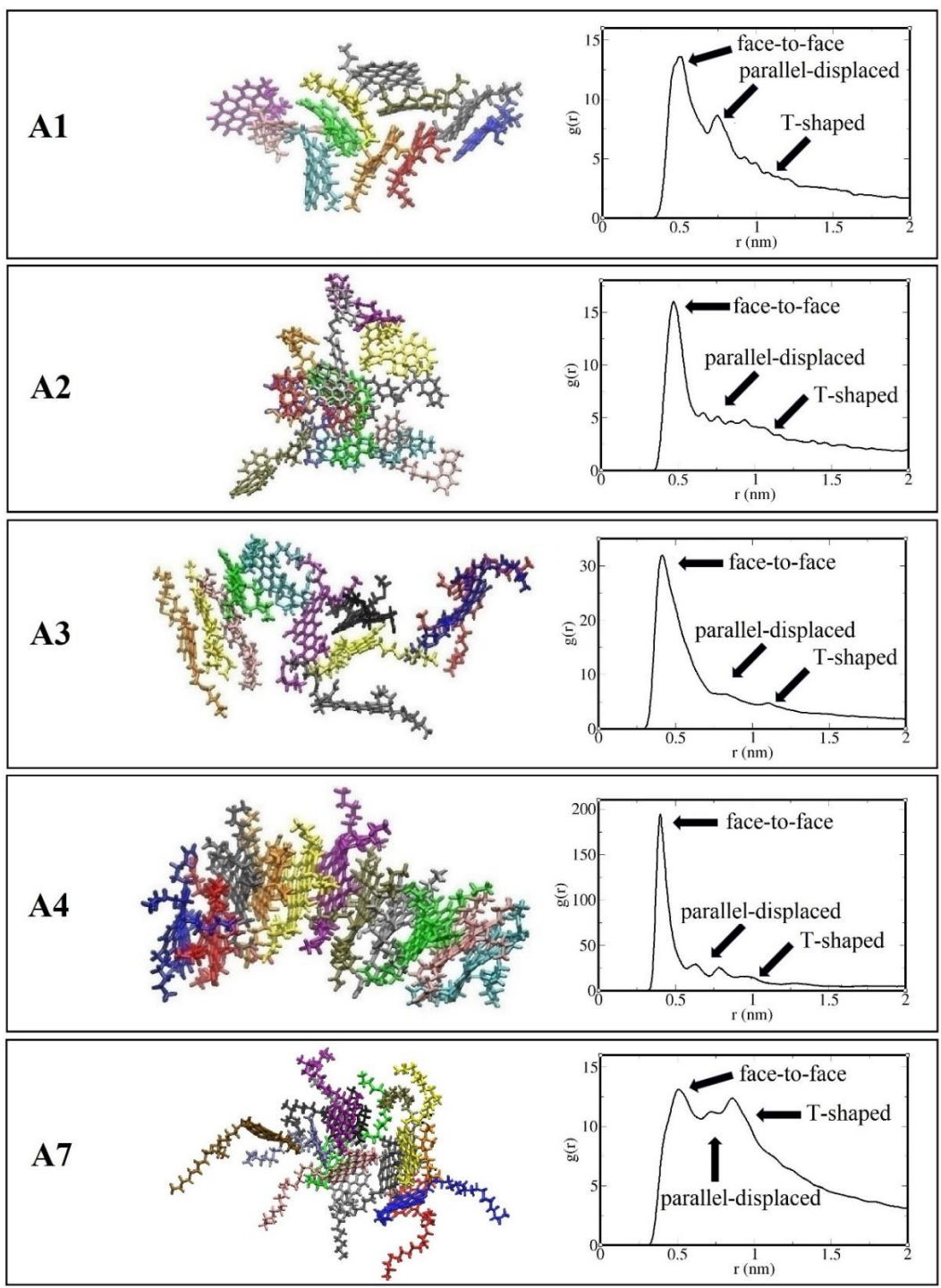

**Figure 3.** Configurational snapshots and RDFs of A1, A2, A3, A4, and A7 aggregates. For clarity purpose, the aggregated asphaltene molecules are shown in various colors. All data are computed at 550 K and 200 bar.

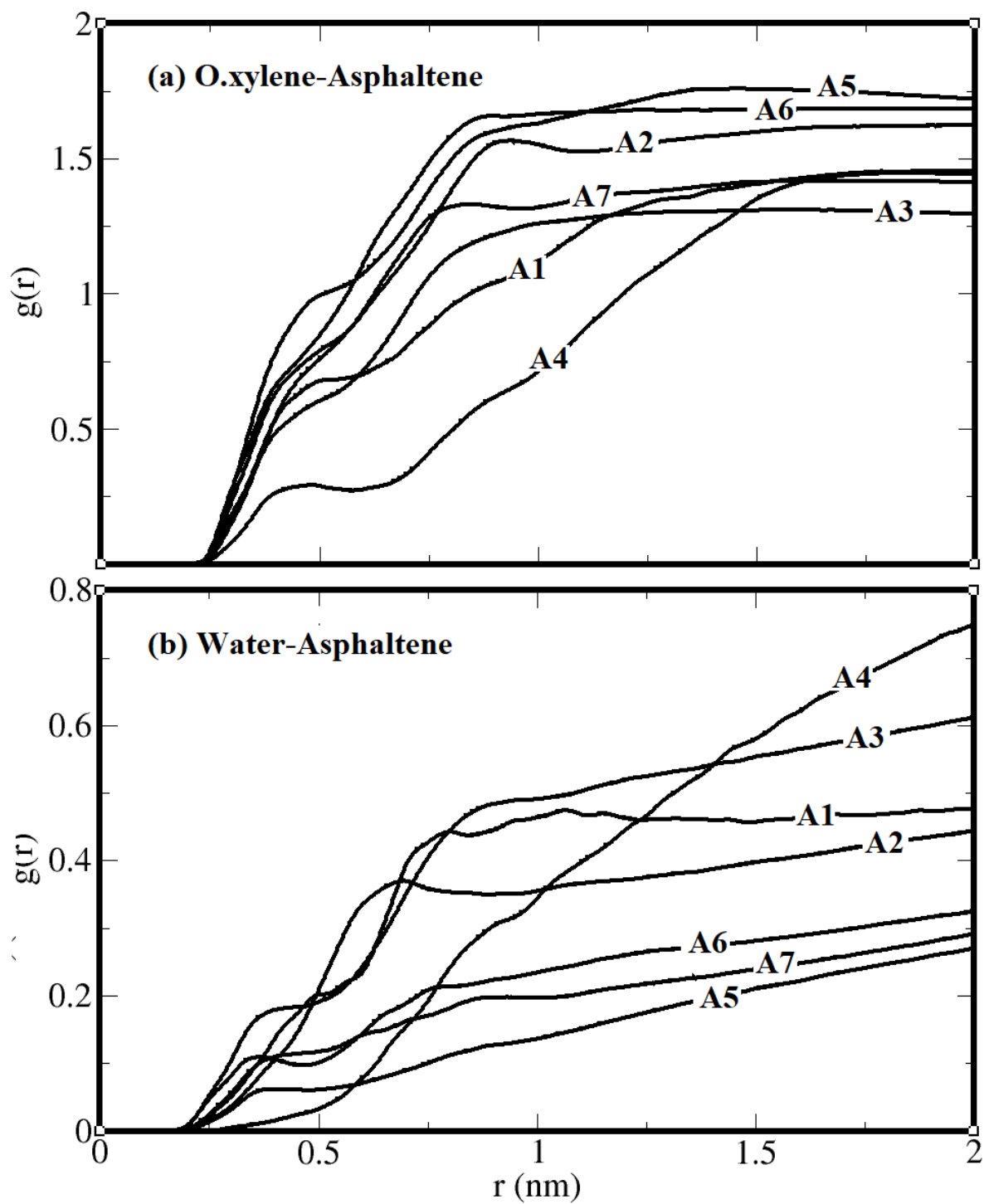

Figure 4. RDFs of o.xylene (a) and water (b) molecules around the aromatic core of asphaltenes. All data are calculated at 550 K and 200 bar.